\documentclass[prd,12pt,tightenlines,nofootinbib]{revtex4}

\usepackage[english]{babel}
\usepackage{graphicx}
\usepackage{psfrag}
\usepackage{amsmath}
\usepackage{amssymb}
\usepackage{bbm}
\usepackage{slashed}
\usepackage{verbatim}
\usepackage[hypertex]{hyperref}

\newcommand{\be}{\begin{equation}}
\newcommand{\ee}{\end{equation}}
\newcommand{\bea}{\begin{eqnarray}}
\newcommand{\eea}{\end{eqnarray}}
\newcommand{\bean}{\begin{eqnarray*}}
\newcommand{\eean}{\end{eqnarray*}}

\renewcommand{\b}{\langle}
\newcommand{\ket}{\rangle}

\newcommand{\irm}{{\rm i}}
\newcommand{\e}{{\rm e}}

\renewcommand{\d}{{\rm d}}
\newcommand{\cl}[1]{{\mathcal #1}}

\renewcommand{\v}[1]{\vec{#1}}

\newcommand{\ds}{\displaystyle}

\newcommand{\bZ}{\mathbb{Z}}

\newcommand{\bC}{\mathbb{C}}
\newcommand{\bR}{\mathbb{R}}

\newcommand{\bH}{\mathbb{H}}

\newcommand{\clH}{\cl{H}}

\newcommand{\clD}{\cl{D}}

\newcommand{\clC}{\cl{C}}

\newcommand{\eq}[1]{(\ref{#1})}
\renewcommand{\sec}[1]{sec.\ \ref{#1}}

\newcommand{\fig}[1]{Fig.\ \ref{#1}}

\newcommand{\tr}{{\rm tr}}

\newcommand{\threevector}[3]
{\left(\begin{array}{c} #1 \\ #2 \\ #3 \end{array}\right)}
\newcommand{\twovector}[2]
{\renewcommand{\arraystretch}{1.5}\left(\begin{array}{c} #1 \\ #2
\end{array}\right)}

\newcommand{\pic}[4]
{
 \begin{figure}
 \begin{center}
 \includegraphics[height=#3]{#4}
 \end{center}
 \caption{\label{#1} #2}
 \end{figure}
}

\newcommand{\qed}{\nobreak \ifvmode \relax \else
      \ifdim\lastskip< 1 em \hskip-\lastskip
      \hskip1.0em plus0em minus0.5em \fi \nobreak
      \vrule height0.75em width0.75em depth0 em\fi}

\newcommand{\xib}{\overline{\xib}}

\begin{document}

\title{A spin foam model for general Lorentzian 4--geometries}

\author{Florian Conrady}
\email{fconrady@perimeterinstitute.ca}
\affiliation{Perimeter Institute for Theoretical Physics, Waterloo, Ontario, Canada}
\author{Jeff Hnybida}
\email{jhnybida@perimeterinstitute.ca}
\affiliation{Perimeter Institute for Theoretical Physics, Waterloo, Ontario, Canada}
\affiliation{Department of Physics, University of Waterloo, Waterloo, Ontario, Canada}

\begin{abstract}
We derive simplicity constraints for the quantization of general Lorentzian 4--geometries.
Our method is based on the correspondence between coherent states and classical bivectors
and the minimization of associated uncertainties. For triangulations with spacelike triangles,
this scheme agrees with the master constraint method of the model by Engle, Pereira, Rovelli and Livine (EPRL).
When it is applied to general triangulations of Lorentzian geometries, we obtain new constraints
that include the EPRL constraints as a special case. They imply a discrete area spectrum
for both spacelike and timelike surfaces. We use these constraints to define a spin foam model
for general Lorentzian 4--geometries.
\end{abstract}

\maketitle

\section{Introduction}
\label{introduction}

What happens if one describes geometry as a degree of freedom of  quantum theory?
Does geometry remain continuous or does it come in quanta?
Are the singularities of classical general relativity resolved?
Loop quantum gravity originates from the attempt to answer such questions\footnote{See \cite{Thiemannbook}, \cite{Rovellibook} and \cite{Perezreview} for reviews.}.
In the Hamiltonian framework, this led to canonical loop quantum
gravity, and, in the path integral picture, it brought forth the notion of
spin foam models \cite{Baezspinfoammodels}.
A central result of both approaches
is the discreteness of the area spectrum: it suggests that there is a minimal unit of area
and that quantum geometry is indeed discrete.

The basic idea behind spin foam models is to divide spacetime
into 4--simplices and to study how the
geometry of these simplices can be quantized. More precisely,
bivectors $B$ are used to describe triangles and constraints
are imposed, so that four bivectors are equivalent to a tetrahedron.
The bivectors and constraints are then translated
in a suitable way to the quantum theory.
The main elements of this procedure were introduced by Barrett
and Crane \cite{BarrettCrane}. In the language of field theory, this corresponds to the transition from
topological BF theory to gravity: the B in BF is constrained to be simple,
so that it becomes the wedge product of two tetrads. For this reason,
the constraints are called simplicity constraints.

In recent years, considerable progress was made in improving and clarifying
this quantization process. The key to this progress were two new developments:
firstly, Engle, Pereira, Rovelli and Livine (EPRL) defined a new model
that resolves certain longstanding problems with the Barrett--Crane model and establishes a link with canonical loop
quantum gravity \cite{EPRL}. The quantization is based on a so--called master constraint,
which is the sum of the squares of all simplicity constraints.
The second important innovation was the coherent state technique introduced by Livine and Speziale
\cite{Livine_New}. It provides a better geometric understanding of quantum states,
and led to the construction of the Freidel--Krasnov (FK) model \cite{Livine_Consistently,FK}. In this model,
simplicity is imposed on expectation values of coherent states.

Both of these developments spurred further results:
The FK model was reexpressed as a path integral with a simple action \cite{CFpathrep}.
The semiclassical limit of the new models was analyzed \cite{CFsemiclassical,BarrettasymptoticsEuclidean,BarrettasymptoticsLorentzian},
likewise the graviton propagator \cite{gravitonpropagatornewmodel}.
It was found that intertwiner states can be understood in geometric terms \cite{CFquantumgeometry}.
Recently, coherent states were constructed for entire 3--geometries \cite{Bianchicoherent,FStwisted}.

The present paper is motivated by two questions. The first question concerns the relation between
the EPRL and coherent state approach. What is the connection between these two lines of thought?
Is there a way to understand the master constraint in terms of coherent states?
In the Riemannian case, we know from explicit comparison that the EPRL and FK model are closely related \cite{CFpathrep}.
To this extent, the coherent state ideas apply also to the EPRL model.
In the Lorentzian case, however, it is not clear how a derivation from coherent states should look like\footnote{A proposal was
made in ref.\ \cite{FK}.}. It could be very useful to have one, since coherent states provide a particularly transparent quantization of
simplicity constraints.

The second motivation for this paper comes from the fact that the EPRL model is only defined for
spacelike area bivectors. That is, it can only describe geometries in which all surfaces are spacelike.
The obvious question is therefore: how can one specify a model that covers realistic Lorentzian geometries,
where bivectors can be both spacelike and timelike?

In this article, we obtain an answer to our first question, and it turns out that we can also resolve the
second question with this knowledge.
What we find is a coherent state method that reproduces the EPRL constraints and applies at the same time
to the cases which were not yet covered in this model.
On the one hand, we recover the EPRL constraints for tetrahedra with timelike normals. Thus,
the method gives the desired coherent state derivation of the EPRL spin foams.
The same scheme, however, works also for tetrahedra with spacelike normals. In this case, it results
in two new sets of constraints that pertain to spacelike and timelike triangles within such tetrahedra.

Our method is inspired by the Riemannian FK model and extends its logic by an additional condition.
We demand that there are quantum states for which
\begin{enumerate}
\item The expectation value of the bivector operator is simple.
\item The uncertainty in the bivector is minimal\footnote{The precise meaning of ``minimal'' is stated in \sec{coherencesimplicitySU(2)} and \sec{coherencesimplicitySU(1,1)}.}.
\end{enumerate}
With this, we enforce the existence of coherent states that correspond to classical simple bivectors.
Such states can only exist in certain irreducible representations of $\mathrm{SL(2,\bC)}$ and its subgroups.
This determines constraints on irreps and we interpret them as the quantum version of the simplicity constraints.

Based on this, we define a new spin foam model that gives a quantization of
tetrahedra with spacelike and timelike normals, and hence a quantization of general Lorentzian geometries.
In addition, a coherent state vertex amplitude is specified.
We should remark that coherent states are not necessary, and, in fact, not used,
when defining the spin foam sum. In this work, the coherent states are only essential
in the derivation of the simplicity constraints.

The paper is organized as follows: in \sec{coherencesimplicitySU(2)} we introduce our coherent state method.
It is applied to tetrahedra with a timelike normal and the constraints of the EPRL model are reproduced.
In \sec{coherencesimplicitySU(1,1)}, we treat tetrahedra with a spacelike normal and obtain two new constraints
that refer to spacelike and timelike triangles respectively. Section \ref{summaryconstraints} summarizes the constraints
for the different cases. In \sec{spinfoammodel}, we use these constraints to define a spin foam model for general Lorentzian
4--geometries.

\subsection*{Conventions}

At the outset, we make a few remarks on notation to avoid confusion due to
 differing conventions in the literature.

Our sign convention for the spacetime metric is $(+,-,-,-)$,
and $(+,-,-)$ for 3d Minkowski spacetime. Moreover, $\epsilon^{0123} = +1$.
The Immirzi parameter is $\gamma$ and we assume that $\gamma > 0$.
Unit normal vectors of tetrahedra are denoted by $U$. The letter $N$ stands for unit normal vectors of triangles.
We have the Hodge dual operator $\star$, and $\star B$ is the area bivector of triangles.

Unitary irreducible representations of $\mathrm{SL(2,\bC)}$
are labelled by pairs $(\rho$,$n)$, where $\rho\in\bR$ and $n\in\bZ_+$.
With regard to irreducible representations of subgroups, we follow the notation in \cite{Davids}:
$\clD_j$ stands for SU(2) irreps. The discrete and continuous series of SU(1,1) are designated
by $\clD^\pm_j$ and $\clC^\epsilon_s$ respectively.
Representation matrices are symbolized by $D^{(\rho,n)}(g)$ in the $\mathrm{SL(2,\bC)}$ case,
and by $D^j(g)$ for subgroups.

\section{Simplicity constraints for the SU(2) reduction}
\label{coherencesimplicitySU(2)}

In this section, we introduce our procedure for deriving the simplicity constraints on representations of the Lorentz group.
We do so by applying it to a situation that has been already treated in the EPRL model \cite{EPRL}.

As explained below, the simplicity constraints split into two main categories, depending on whether
normal vectors $U$ of tetrahedra are timelike or spacelike. The case considered in \cite{EPRL} is the one
for timelike $U$. At the quantum level, this choice of a timelike normal
is reflected by the fact that unitary irreps of $\mathrm{SL(2,\bC)}$
are analyzed in terms of irreps of the subgroup SU(2). For this case, we will find that our method produces the same
simplicity constraints as the master constraint employed in the EPRL paper.

Encouraged by this agreement, we will then proceed to section \ref{coherencesimplicitySU(1,1)},
where we apply our technique to the case when $U$ is spacelike. Then, irreps of
$\mathrm{SL(2,\bC)}$ will be decomposed into irreps of SU(1,1) and we will obtain a
new set of simplicity constraints.

\subsection{Classcial variables}

Our starting point is the SO(1,3) bivector
\be
J = B + \frac{1}{\gamma} \star\!B\,,
\label{Jbivector}
\ee
which is used when defining gravity as a constrained BF theory with Immirzi parameter $\gamma$.
The idea is to constrain $B$ in such a way that $B$ becomes $B = \star(E\wedge E)$, where $E$
is a co--tetrad. Under this constraint, the action
\be
S = \int J\wedge F = \int \left(B\wedge F + \frac{1}{\gamma} \star B\wedge F\right)
\ee
reduces to the Hilbert--Palatini action with an Immirzi term.

Like in \cite{EPRL}, we denote the total bivector by the letter $J$.
This choice is convenient, since $J$ will be closely related to the generators
of the Lorentz group. After quantization, the classical bivector $J^{IJ}$ will
arise from expectation values of SO(1,3) generators $J^{IJ}$.

At the classical level, the derivation of the simplicity constraints proceeds as follows:
first we state the simplicity constraints for $B$, which ensure that $B = \star(E\wedge E)$.
These constraints involve a vector $U$ which is normal to the dual bivector $\star B$.
Since $U$ can be either timelike or spacelike, one obtains two classes of simplicity constraints.
In the present section, we treat the case where $U$ is timelike, as in the EPRL model.
These constraints are then expressed as constraints on $J$---ready to be translated to the
quantum theory.

In full generality, the simplicity constraint reads
\be
U\cdot \star B = 0\,,
\label{simplicityconstraint}
\ee
where $U$ is a Lorentz vector of unit norm, i.e.\ $U^2 = \pm 1$.
This constraint implies that the dual $\star B$ is simple and of the form\footnote{$V\wedge W$ stands for the bivector $(V\wedge W)^{IJ} = V^I W^J - W^I V^J$.}
\be
\star\!B = E_1\wedge E_2\,,
\label{simple*B}
\ee
where $E_1$ and $E_2$ are two 4--vectors orthogonal to $U$ (for a derivation see e.g.\ \cite{FK}).
Moreover, $B$ is simple and given by
\be
B = A\, U\wedge N\,,
\label{simpleB}
\ee
where $N$ is a unit norm 4--vector such that $U\cdot N = 0$ and $N\cdot E_1 = N\cdot E_2 = 0$. The coefficient $A$ in \eq{simpleB} is equal to the area
\be
A = \sqrt{\left|E^2_1 E^2_2 - (E_1\cdot E_2)^2\right|}
\ee
of the parallelogram spanned by $E_1$ and $E_2$.

Equations \eq{simple*B} and \eq{simpleB} elucidate the geometric meaning of the 4--vectors $U$ and $N$.
$E_1$ and $E_2$ correspond to the tetrad and in a discrete setting they can be regarded as the two edges of a triangle.
The dual $\star B$ is the so--called area bivector of this triangle, and it is orthogonal to both $U$ and $N$.
To obtain a tetrahedron, one starts from four bivectors $B_a$, $a = 1,\ldots,4,$ and imposes the simplicity constraint \eq{simplicityconstraint} on each of them:
\be
U\cdot \star B_a = 0\,,\quad a = 1,\ldots,4\,.
\label{simplicityfourbivectors}
\ee
This means that all four bivectors are simple and that they span a 3d subspace orthogonal to $U$.
When the closure constraint
\be
\sum_{a=1}^4 B_a = 0\,,
\ee
is supplemented, it follows that the bivectors $B_a$ are equivalent to a tetrahedron.
The vector $U$ is the normal of this tetrahedron, the four vectors $N_a$ (associated to the bivectors $B_a$) are the normals to the four triangles,
and the $E$'s are the edges of the triangles. In this way, geometry results from constraints on bivectors.
In spin foam models the analog of \eq{simplicityfourbivectors} is imposed on representations of the gauge group,
while the closure constraint arises dynamically.

Let us assume now that $U$ is timelike and gauge--fixed to $U = (1,0,0,0)$. Then, the simplicity constraint becomes $(\star B)^{0i} = 0$
and $\star B$ has to be spacelike. Next we express this constraint in terms of the bivector $J$. Solving for $J$ in \eq{Jbivector} gives
\be
B = \frac{\gamma^2}{\gamma^2 + 1}\left(J - \frac{1}{\gamma} \star J\right)\,.
\label{BintermsofJ}
\ee
It follows that the gauge--fixed simplicity constraint is equivalent to
\be
J^i + \frac{1}{\gamma} K^i = 0
\label{simplicitySU(2)}
\ee
where we use the usual definitions
\be
J^i = \frac{1}{2}\,\epsilon^{0i}{}_{jk} J^{jk}\qquad\mbox{and}\qquad K^i = J^{0i}\,.
\ee
Eq.\ \eq{simplicitySU(2)} will be the central equation for the derivation of the simplicity constraints in the quantum theory.

It is also useful to express the normal vector $N$ in terms of $J^i$ and $K^i$.
On the one hand, we have that
\be
B^{ij} = 0\,,\qquad B^{0i} = A\,N^i\,.
\ee
On the other hand,
\bea
B^{0i} &=& \frac{\gamma^2}{\gamma^2 + 1}\left(J^{0i} - \frac{1}{2\gamma} \epsilon^{0i}{}_{jk} J^{jk}\right) \\
&=& \frac{\gamma^2}{\gamma^2 + 1}\left(K^i - \frac{1}{\gamma} J^i\right)
\eea
Using the simplicity constraint \eq{simplicitySU(2)} this yields
\be
A\,N^i = -\gamma J^i
\label{relationNJ}
\ee
Therefore, $N = (0,\v{N})$, where $A \v{N} = -\gamma \v{J}$.
Observe also that the vector $\v{N}$ is a point in the 2--sphere
\be
S^2 \simeq \mathrm{SU(2)} / \mathrm{U(1)}\,.
\ee
and hence a point in the coadjoint orbit (or phase space) of spin $j = 1$.

\subsection{Quantum states}

Next we will describe our way of translating the simplicity constraint to quantum states.
As in the EPRL model, we will not do this in a manifestly covariant way.
We will start from eq.\ \eq{simplicitySU(2)}, which is the simplicity constraint
after gauge--fixing $U$ to $(1,0,0,0)$.

The quantum analog of the bivectors are states in unitary irreducible representations of $\mathrm{SL(2,\bC)}$.
The latter are labelled by  pairs $(\rho,n)$, where $\rho\in\bR$ and $n\in\bZ_+$.
Since $U = (1,0,0,0)$ singles out SU(2) as a little group of $\mathrm{SL(2,\bC)}$,
it is convenient to express everything in an ``SU(2) friendly'' way.
This can be done by using the decomposition of $\mathrm{SL(2,\bC)}$ irreps into SU(2) irreps: namely,
\be
\clH_{(\rho,n)} \simeq \bigoplus\limits_{j = n/2}^{\infty} \clD_j
\ee
where $\clH_{(\rho,n)}$ denotes the Hilbert space of the irrep $(\rho,n)$ and
$\clD_j$ stands for the spin $j$ irrep of SU(2) (see e.g.\ \cite{Ruhl}). The corresponding completeness relation reads
\be
\mathbbm{1}_{(\rho,n)} = \sum\limits_{j = n/2}^\infty \sum_{m=-j}^j \left|\Psi_{j\, m}\right\ket \left\b\Psi_{j\, m}\right|\,.
\label{completenessrelationSU(2)}
\ee
The states $\left|\Psi_{j\, m}\right\ket$, $m = -j,\ldots, j$, span a subspace of $\clH_{(\rho,n)}$ that is isomorphic
to $\clD_j$, so we identify them with states $|j\, m\ket$ of $\clD_j$.

\setlength{\jot}{0.4cm}
Our recipe for quantizing the simplicity constraints is formulated as follows.
We require the existence of quantum states for which the expectation value of the
bivector $J^{IJ}$ satisfies the simplicity constraints. Moreover, the quantum uncertainty in this
bivector should be small.
Here, we work with a gauge--fixing, so the components of $J^{IJ}$ are organized in terms
of $\v{J}$ and $\v{K}$. Let us define associated lengths $|\v{J}| \equiv |\b \v{J}\ket|$ and
$|\v{K}| \equiv |\b \v{K}\ket|$,
where $\b\;\ket$ stands for the expectation value w.r.t.\ the quantum state.
We then demand that
\bea
&& \frac{\Delta J}{|\v{J}|} = O\left(\frac{1}{\sqrt{|\v{J}|}}\right)\,,
\label{uncertaintyJ} \\
&&
\b \v{J} \ket + \frac{1}{\gamma} \b \v{K}\ket = O(1)\,,
\label{quantumsimplicity} \\
&&
\frac{\Delta K}{|\v{K}|} = O\left(\frac{1}{\sqrt{|\v{K}|}}\right)\,.
\label{uncertaintyK}
\eea
Through these three conditions we establish a correspondence
between classical variables and semiclassical states:
the first requirement says that the states
should be peaked around classical values of $\v{J}$.
The second condition states that their expectation values
fulfill the simplicity constraint.
The last point adds that the states should
not only be peaked in $\v{J}$, but also in the remaining components
$\v{K}$.

The first condition is easily met by using SU(2) coherent states \cite{Perelomov}: these states have the form
\be
|j\,g\ket \equiv D^j(g) |j\,j\ket\,.
\ee
and arise from SU(2) rotations of the ``reference'' coherent state $|j\, j\ket$. From such states we get
\be
\frac{\Delta J}{|\v{J}|} = \frac{\sqrt{j}}{j} = \frac{1}{\sqrt{j}} = O\left(\frac{1}{\sqrt{|\v{J}|}}\right)\,.
\ee

\setlength{\jot}{0.3cm}
Equation \eq{quantumsimplicity} is more subtle, as it involves $\mathrm{SL(2,\bC)}$ generators outside of SU(2).
Since both $\v{J}$ and $\v{K}$ transform as vectors under SU(2), it is sufficient to impose \eq{quantumsimplicity} on the reference state $|j\, j\ket$.
If it is satisfied for $|j\, j \ket$, then it will be true for all coherent states $|j\, g\ket$.
Therefore, we require\footnote{For the next couple of lines we omit the order symbol $O(1)$.}
\be
\b j\,j | \v{J} | j\,j\ket = -\frac{1}{\gamma} \b j\,j | \v{K} | j\,j\ket\,.
\label{simplicityreferencestate}
\ee
It is clear from commutation relations that $J^1$, $J^2$, $K^1$ and $K^2$ change the eigenvalue of $J^3$,
so their expectation values will be zero. We therefore only need to consider $J^3$ and $K^3$. The action of $K^3$ is given by\footnote{See, for instance, \cite{Carmeli}.}
\bea
K^3 |j\, m\ket &=&
- \sqrt{(j+m+1)(j-m+1)}\, C_{j+1}\, |j+1\, m\ket \nonumber \\
&& {}- m A_j\, |j m\ket \nonumber \\
&& {}+ \sqrt{(j-m)(j+m)}\, C_j\, |j-1\, m\ket
\,,
\eea
where
\be
A_j = \frac{\rho\, n}{4j(j+1)}\,,
\ee
and
\be
C_j = \frac{\irm}{j} \sqrt{\frac{(j^2-\frac{n^2}{4})(j^2+\frac{\rho^2}{4})}{4j^2 - 1}}\,.
\ee
Hence eq.\ \eq{simplicityreferencestate} leads to
\be
j = -\frac{1}{\gamma}(-j A_j)\qquad\mbox{or}\qquad \gamma = A_j = \frac{\rho\, n}{4j(j+1)}\,.
\label{simplicityconditionSU(2)}
\ee

In order to deal with the variance in $\v{K}$, we recall the Casimirs of $\mathrm{SL(2,\bC)}$:
\bea
C_1 &=& 2\left(\v{J}^2 - \v{K}^2\right) = \frac{1}{2} (n^2 - \rho^2 - 4) \\
C_2 &=& -4 \v{J}\cdot\v{K} = n\rho
\eea
As a result, one gets
\bea
\left(\Delta K\right)^2 &=& \b \v{K}^2 \ket - \b \v{K} \ket^2  \\
&=& \b \v{J}^2 - \frac{1}{4} (n^2 - \rho^2 - 4) \ket - \b \v{K} \ket^2 \\
&=& j(j+1) - \frac{1}{4} (n^2 - \rho^2 - 4) - j^2 A^{2}_j\,.
\eea
By inserting the simplicity constraint $A_j = \gamma$, we obtain furthermore
\bea
\left(\Delta K\right)^2 &=& \frac{1}{4 \gamma}{\rho\, n} - \frac{1}{4} (n^2 - \rho^2 - 4) - j^2 \gamma^2  \\
&=&
\frac{1}{4 \gamma}{\rho\, n} - \frac{1}{4} (n^2 - \rho^2) - \gamma^2 j(j+1) + \gamma^2 j  + 1  \\
&=&
\frac{1}{4 \gamma}{\rho\, n} - \frac{1}{4} (n^2 - \rho^2) - \frac{\gamma}{4} \rho\, n + \gamma^2 j + 1  \\
&=&
\frac{1}{4} \Big(\rho - \gamma n \Big) \Big(\rho + \frac{n}{\gamma} \Big) + \gamma^2 j  + 1\,.
\label{varianceKintermsofrhon}
\eea
If one sets
\be
\rho = \gamma n
\label{solution1}
\ee
or
\be
\rho = -\frac{n}{\gamma}\,,
\label{solution2}
\ee
the first term vanishes, and
\be
\frac{\Delta K}{|\v{K}|} = \frac{\sqrt{\gamma^2 j + 1}}{\gamma j} = O\left(\frac{1}{\sqrt{|\v{K}|}}\right)\,.
\ee
However, when plugging back \eq{solution1} into the simplicity constraint \eq{simplicityconditionSU(2)}, we obtain
\be
n^2 = 4 j (j+1)\,.
\ee
Since this cannot be solved for generic values of $n$ and $j$, we proceed like in \cite{EPRL} and adopt the approximate solution
\be
j = n/2\,.
\label{approximatesolution}
\ee
One can check that \eq{solution1} and \eq{approximatesolution} fulfill conditions \eq{quantumsimplicity} and \eq{uncertaintyK}.
When \eq{solution2} is inserted into \eq{simplicityconditionSU(2)}, on the other hand, we get
\be
n^2 = - 4\gamma^2 j(j+1)\,.
\ee
This has no solution. Thus, our final result are the constraints $\rho = \gamma n$ and $j = n/2$, which are the same constraints as
in the EPRL model!

The area spectrum can be derived by squaring the classical equation \eq{relationNJ} and setting the right--hand side equal to the expectation
value of the coherent state. This gives us the quantum area
\be
A = \gamma \sqrt{\b \v{J}^2\ket} = \gamma \sqrt{j(j+1)}\,.
\label{quantumareaSU(2)}
\ee

Up to this point, we have just used a new perspective to obtain something that was already known, i.e.\ the constraints of the EPRL model.
This shows that the EPRL master constraint is equivalent to a set of semiclassical constraints,
namely to the requirement that there exist quantum states that are peaked in
$\v{J}$ and $\v{K}$, and that their expectation values satisfy the classical simplicity constraint.

In the next section, we will apply our procedure to derive something new:
we will provide a prescription for spacelike $U$, and hence for bivectors $\star B$ that can be both spacelike and timelike.

\section{Simplicity constraints for the SU(1,1) reduction}
\label{coherencesimplicitySU(1,1)}

\subsection{Classical variables}

\renewcommand{\arraystretch}{1.5}
In this section, the normal vector $U$ is assumed to be spacelike, and we gauge--fix it to $U = (0,0,0,1)$.
Then, the classical simplicity constraint is $(\star B)^{3i} = 0$, where $i = 0, 1, 2$. A short calculation
shows that this is equivalent to
\bea
K^1 - \frac{1}{\gamma} J^1 = 0\,, \\
K^2 - \frac{1}{\gamma} J^2 = 0\,, \\
J^3 + \frac{1}{\gamma} K^3 = 0\,.
\eea
By defining the quantities
\[
\begin{array}{lll}
F^0 = J^3\,, & F^1 = K^1\,, & F^2 = K^2\,, \\
G^0 = K^3\,, & G^1 = -J^1\,, & G^2 = -J^2\,,
\end{array}
\]
one can write these constraints in the more symmetric form
\be
F^i + \frac{1}{\gamma} G^i = 0\,,\qquad i = 0, 1, 2\,.
\label{simplicitySU(1,1)}
\ee
Here and below we use the indices $i, j$ to denote vectors in 3--dimensional Minkowski spacetime,
in the same way that indices $i, j$ were used for vectors of 3--dimensional Euclidean space in the previous section.
Inspection of the commutation relations reveals, in fact, that $F$ and $G$ transform like 3d Minkowski vectors
under SU(1,1) \cite{Mukundahomogeneous}:
\bea
&& [F^i,F^j] = \irm\,C^{ij}{}_k F^k\,, \\
&& [F^i,G^j] = \irm\,C^{ij}{}_k G^k\,,
\eea
where
\be
C^{01}{}_2 = -C^{10}{}_2 = C^{20}{}_1 = -C^{02}{}_1 = C^{21}{}_0 = -C^{12}{}_0 = 1\,.
\ee
Thus, eq.\ \eq{simplicitySU(1,1)} has a similar structure as eq.\ \eq{simplicitySU(2)}:
it relates two vectors that have the same transformation property under the little group,
and one of the vectors is the generator of the little group.

As before, we determine the relation between the vector $N$ and the generators.
On the one hand,
\be
B^{ij} = 0\,,\qquad B^{3i} = A\,N^i\,,\qquad\mbox{where $i,j = 0, 1, 2$}\,.
\ee
On the other hand, we also have
\be
B^{3i} = \frac{\gamma^2}{\gamma^2 + 1}\left(J^{3i} - \frac{1}{2\gamma} \epsilon^{3i}{}_{jk} J^{jk}\right)
\ee
For $i = 0$, this gives
\bea
B^{30} &=& \frac{\gamma^2}{\gamma^2 + 1}\left(J^{30} - \frac{1}{\gamma} \epsilon^{30}{}_{12} J^{12}\right) \\
&=& \frac{\gamma^2}{\gamma^2 + 1}\left(-K^3 + \frac{1}{\gamma} J^3\right)
\eea
On account of the simplicity constraint this reduces to
\be
A N^0 = \gamma J^3\,.
\ee
Similarly, we obtain
\be
A N^1 = \gamma K^2\qquad\mbox{and}\qquad A N^2 = -\gamma K^1
\ee
for the cases $i = 1$ and $i = 2$.
Altogether we get
\be
A \threevector{N^0}{N^1}{N^2} = \gamma \threevector{J^3}{K^2}{-K^1} = \gamma \threevector{F^0}{F^2}{-F^1}\,.
\label{relationNF}
\ee
Like in the SU(2) case, the vector $\v{N} = (N^0,N^1,N^2)$ is related to quotient spaces of SU(1,1):
timelike vectors $\v{N}$ coordinatize the two--sheeted hyperboloid
\be
\bH_+ \cup \bH_- \,,\qquad \bH_\pm = \{\,\v{N}\;|\; \v{N}^2 = 1\,, N^0 \gtrless 0\,\}\,,
\ee
and each sheet is isomorphic to the quotient $\mathrm{SU(1,1)} / \mathrm{U(1)}$.
Spacelike $\v{N}$ parametrize the spacelike single--sheeted hyperboloid $\bH_{\mathrm{sp}} = \{\v{N}\,|\, \v{N}^2 = -1\}$.
This hyperboloid is isomorphic to the quotient $\mathrm{SU(1,1)} / (\mathrm{G_1}\otimes\bZ_2)$,
where $G_1$ is the one--parameter subgroup of SU(1,1) generated by $K_1$ \cite{Lindblad}.
Appendix \ref{parametrization} describes the details of these isomorphisms.

\subsection{Quantum states}

\setlength{\jot}{0.3cm}
Since we have set $U$ equal to $(0,0,0,1)$, the little group is SU(1,1).
Accordingly, we should work with unitary irreducible representations of SU(1,1).
These come in a discrete and a continuous series. In both cases, the irreps
can be built from eigenstates $|j\,m\ket$ of $J^3$, satisfying
\bea
\b j\, m | j\, m'\ket &=& \delta_{mm'}\,, \\
J^3\, |j\, m\ket &=& m |j\, m\ket\,.
\eea
Combinations of $K^1$ and $K^2$ act as raising and lowering operators.
The Casimir is given by $Q = (J^3)^2 - (K^1)^2 - (K^2)^2$.

In irreps of the discrete series, one has
\be
Q\, |j\, m\ket = j(j-1) |j\, m\ket\,,\qquad\mbox{where $j$ = $\frac{1}{2}$, $1$, $\frac{3}{2}$, \ldots}
\ee
The eigenvalue $m$ can take the values
\be
m = j,\; j+1,\; j+2,\; \ldots\qquad\mbox{or}\qquad m = -j,\; -j-1,\; -j-2,\; \ldots
\ee
We use the same notation as in \cite{Davids} and denote the irrep consisting of states $|j\, m\ket$ with $m \gtrless 0$
by $\clD^\pm_j$. The fact that $Q = j(j-1)$ is positive for $j \ge 3/2$ suggests that the discrete series contains
coherent states corresponding to timelike 3--vectors.

For the continuous series,
\be
Q\, |j\, m\ket = j(j+1) |j\, m\ket\,, \qquad\mbox{where $j = -\frac{1}{2} + \irm s$,\quad $0 < s < \infty$,}
\ee
and
\be
m = 0,\, \pm 1,\, \pm 2,\, \ldots\qquad\mbox{or}\qquad m = \pm\frac{1}{2},\, \pm\frac{3}{2},\, \ldots
\ee
Irreps of this series are denoted by $\clC^\epsilon_s$. The label $\epsilon = 0, \frac{1}{2}$ designates the irreps with integer $m$ and half--integer $m$
respectively. In this case, the Casimir $Q = j(j+1) = -s^2 - \frac{1}{4}$ is always negative, so we expect
coherent states to be associated to spacelike 3--vectors.

\renewcommand{\arraystretch}{2.5}
Similarly as for SU(2), the $\mathrm{SL(2,\bC)}$ irrep $(\rho,n)$ can be expanded in an SU(1,1)--adapted basis (see \cite{Mukundahomogeneous,Ruhl} and also \cite{Bargmann}).
The resulting completeness relation involves states $\left|\Psi^\pm_{j\, m}\right\ket$ and $|\Psi^{(\alpha)}_{s\, m}\ket$, $\alpha = 1,2$,
that correspond to states $|j\, m\ket$ in the discrete and continuous series respectively:
\bea
\mathbbm{1}_{(\rho,n)} &=& \sum\limits_{j > 1/2}^{n/2} \sum_{m=j}^\infty \left|\Psi^+_{j\, m}\right\ket \left\b\Psi^+_{j\, m}\right| \nonumber \\
&+& \int\limits_0^\infty \d s\;\mu_\epsilon(s) \sum\limits_{\pm m = \epsilon}^\infty \left|\Psi^{(1)}_{s\, m}\right\ket \left\b\Psi^{(1)}_{s\, m}\right| \nonumber \\
&+& \sum\limits_{j > 1/2}^{n/2} \sum_{-m=j}^\infty \left|\Psi^-_{j\, m}\right\ket \left\b\Psi^-_{j\, m}\right| \nonumber \\
&+& \int\limits_0^\infty \d s\;\mu_\epsilon(s) \sum\limits_{\pm m = \epsilon}^\infty \left|\Psi^{(2)}_{s\, m}\right\ket \left\b\Psi^{(2)}_{s\, m}\right|
\label{completenessrelationSU(1,1)}
\eea
The sum over $j$ extends over values such that $j - n/2$ is integral. Moreover, $\epsilon$ has a value such that $\epsilon - n/2$ is an integer. Note that the irrep $j = 1/2$ does not appear in the expansion.
The measure factors are given by
\be
\mu_\epsilon(s) = \left\{
\begin{array}{ll}
\ds 2 s\tanh(\pi s)\,, & \epsilon = 0\,, \\
\ds 2 s\coth(\pi s)\,, & \epsilon = 1/2\,.
\end{array}
\right.
\ee
When $\mathrm{SL(2,\bC)}$ is restricted to SU(1,1), the states $\left|\Psi^\pm_{j\, m}\right\ket$ furnish irreducible representations
that are isomorphic to those of the discrete series:
\be
\big\b\Psi^{\pm}_{j\, m'} \big|\Psi^\pm_{j\, m}\big\ket = \delta_{m'm}
\ee
\be
\big\b\Psi^{\pm}_{j\, m'} \big| D^{(\rho,n)}(g) \big|\Psi^\pm_{j\, m}\big\ket = \b j\, m'| D^j(g) |j\, m\ket\quad\mbox{for $g\in\mathrm{SU(1,1)}$.}
\label{reducedrepresentation}
\ee
We therefore identify $\left|\Psi^\pm_{j\, m}\right\ket$ with $|j\,m\ket$ in $\clD^\pm_j$.

\renewcommand{\arraystretch}{3.5}
With regard to the continuous series, the situation is more subtle. Firstly, the continuous series states $\left|\Psi^{(\alpha)}_{s\, m}\right\ket$ appear twice,
which is indicated by the index $\alpha = 1,2$. Moreover, these states are not normalizable:
\be
\big\b\Psi^{(\alpha')}_{s'\, m'}\big|\Psi^{(\alpha)}_{s\, m}\big\ket = \frac{\delta(s'-s)}{\mu_\epsilon(s)}\,\delta_{\alpha'\alpha}\,\delta_{m'm}\,,
\label{orthogonalitycontinuousseries}
\ee
As a result, the analog of eq.\ \eq{reducedrepresentation} requires an integration over $s$:
\be
\int\limits_0^\infty \d s'\;\mu_\epsilon(s') \big\b\Psi^{(\alpha)}_{s'\, m'}\big| D^{(\rho,n)}(g)\big|\Psi^{(\alpha)}_{s\, m}\big\ket
= \b j\, m'| D^j(g) |j\, m\ket\,,\qquad j = -\frac{1}{2} + \irm s\,.
\ee
With this qualification in mind, we can say that
\be
\clH_{(\rho,n)} \quad\simeq\quad
\left(\bigoplus\limits_{j > 1/2}^{n/2} \clD^+_j \oplus \int\limits_0^\infty \d s\; \clC^\epsilon_s\right)
\oplus
\left(\bigoplus\limits_{j > 1/2}^{n/2} \clD^-_j \oplus \int\limits_0^\infty \d s\; \clC^\epsilon_s\right)\,.
\label{decompositionSU(1,1)}
\ee
The proof of this decomposition proceeds similarly as for SU(2) \cite{Ruhl}.
By homogeneity, the representation of $\mathrm{SL(2,\bC)}$ on functions of $\bC^2$ reduces to a representation
on pairs of functions $(\varphi_1,\varphi_2)$ of SU(1,1).
Such functions can be expanded into matrix elements of SU(1,1) \cite{Bargmann}, in analogy to the Peter--Weyl theorem for
compact groups. Covariance properties require that certain irreps do not appear in this
decomposition---in the same way that spins $j < n/2$ do not appear in the decomposition into SU(2) irreps.
Thus, one obtains the first two lines in \eq{completenessrelationSU(1,1)} from the first component of the pair, and the last two lines
from the second component.
Explicitly, the states $\left|\Psi^\pm_{j\, m}\right\ket$ and $\left|\Psi^{(\alpha)}_{s\, m}\right\ket$ are
given by
\[
\begin{array}{l@{\qquad}l}
\Psi^+_{j\, m}(g) = \sqrt{2j-1} \twovector{D^j_{n/2, m}(g)}{0}\,, &
\Psi^{(1)}_{s\, m}(g) = \twovector{D^{-1/2+\irm s}_{n/2, m}(g)}{0}\,, \\
\Psi^-_{j\, m}(g) = \sqrt{2j-1} \twovector{0}{D^j_{-n/2, m}(g)}\,, &
\Psi^{(2)}_{s\, m}(g) = \twovector{0}{D^{-1/2+\irm s}_{-n/2, m}(g)}\,,
\end{array}
\]
where $g\in \mathrm{SU(1,1)}$.

\subsection{Constraints for the discrete series}

Let us begin by deriving the simplicity constraints of the discrete series.
The main difference to the SU(2) case is that we are now dealing with Minkowksi 3--vectors.
How can one generalize the notions of minimal uncertainty and coherent states to a relativistic
setting?

In a relativistic theory physical quantities are Lorentz invariant.
Thus, it seems natural to define the uncertainty in the Minkowski vector $F$ by
\bea
\left(\Delta F\right)^2 &=& \Big\b (F - \b F\ket)^i (F - \b F\ket)_i \Big\ket \\
&=& \b F^i F_i \ket - \b F^i \ket \b F_i \ket \label{Minkowskivariance} \\
&=& j(j-1) - m^2 = j^2 - j - m^2\,. \label{variancediscrete}
\eea
Since $j > 0$, and $|m| \ge j$, we see that $(\Delta F)^2$ is always negative.

Our semiclassical conditions  can be easily adapted to this new situation: to accommodate for minus signs,
we define $\Delta F \equiv \sqrt{|\left(\Delta F\right)^2|}$ and $|\v{F}| \equiv \sqrt{|\b \v{F}\ket^2|}$,
and demand that
\bea
&& \frac{\Delta F}{|\v{F}|} = O\left(\frac{1}{\sqrt{|\v{F}|}}\right)\,,
\label{uncertaintyJdiscrete} \\
&&
\b \v{F} \ket + \frac{1}{\gamma} \b \v{G}\ket = O(1)\,,
\label{quantumsimplicitydiscrete} \\
&&
\frac{\Delta G}{|\v{G}|} = O\left(\frac{1}{\sqrt{|\v{G}|}}\right)\,.
\label{uncertaintyKdiscrete}
\eea
The first equation can be solved by choosing states with $m = \pm j$. More generally,
we can use any state of the form
\bea
|j\,g\ket_+ &\equiv& D^j(g) |j\,j\ket\,, \\
|j\,g\ket_- &\equiv& D^j(g) |j -\!j\ket\,,
\eea
where $g\in\mathrm{SU(1,1)}$.
The states $|j\, g\ket_+$ are the coherent states defined by Perelomov for SU(1,1) \cite{Perelomov}.
Observe that the expectation value of $F$ w.r.t.\ $|j\,\mathbbm{1}\ket_+$ gives the vector $\v{N} = (1,0,0)$, while $|j\,\mathbbm{1}\ket_-$ produces $\v{N} = (-1,0,0)$.

These states exhibit a nice relation with the hyperboloids $\bH_+$ and $\bH_-$ mentioned above:
every element $g\in \mathrm{SU(1,1)}$ can be written as
\be
g = g_0 h\,,
\ee
where $h\in \mathrm{U(1)}$. We see therefore that a representative $g_0$ of the coset $\bH_\pm \simeq \mathrm{SU(1,1)}/\mathrm{U(1)}$ is sufficient to determine
the coherent states $|j\, g\ket_\pm$ up to a phase. Moreover, in a completeness relation the coherent states appear both as a bra and a ket,
so that the phase cancels. Thus, it is sufficient to consider states
\be
|j\,\v{N}\ket \equiv |j\, g(\v{N})\ket_\pm\,,
\ee
where $\v{N}\in \bH_\pm$ and $g(\v{N})$ is a representative in the coset defined by $\v{N}$ (see appendix \ref{parametrization}).

The treatment of condition \eq{quantumsimplicitydiscrete} and \eq{uncertaintyKdiscrete} is in many ways analogous to what we did in the previous section.
Consider first the simplicity constraint \eq{quantumsimplicitydiscrete}: since $F$ and $G$ transform in the same way under SU(1,1), it suffices to compute
 the expectation values of the reference coherent states $|j -\!\!j\ket$ and $|j\, j\ket$.
Due to the commutation relations, $K^1$, $K^2$, $J^1$ and $J^2$ change the eigenvalue of $J^3$,
so their expectation values will be zero. It remains to evaluate the expectation values of $J^3$ and $K^3$. According to ref.\ \cite{Mukundahomogeneous} the action of $K^3$ is given by
\be
K^3\, |j\, m\ket =
(\ldots) |j+1\, m\ket
- m \tilde{A}_j |j\, m\ket
+ (\ldots) |j-1\, m\ket\,,\qquad \tilde{A} = \frac{\rho\, n}{4j(j-1)}\,.
\ee
$(\ldots)$ stands for factors that we do not need below, since we are only interested in expectation values.
Therefore, for both $|j\, j\ket$ and $|j -\!j\ket$, the simplicity constraint leads to
\be
j + \frac{1}{\gamma} (-j \tilde{A}_j) = 0
\ee
or
\be
\gamma = \tilde{A}_j = \frac{\rho\, n}{4j(j-1)}\,,
\ee
in analogy to the SU(2) case.
Consider next the variance of $G$:
\be
(\Delta G)^2 \equiv (\Delta G)^i (\Delta G)_i = \b G^i G_i\ket - \b G^i\ket \b G_i\ket
\ee
Noting that
\be
\frac{1}{2} C_1 = \v{J}^2 - \v{K}^2 = Q - \v{G}^2
\ee
we get
\be
\b \v{G}^2\ket = j(j-1) - \frac{1}{4}\left(n^2 - \rho^2 - 4\right)
\ee
and for a state $|j\, m\ket$
\be
(\Delta G)^2 = j(j-1) - \frac{1}{4}\left(n^2 - \rho^2 - 4\right) - (m A_j)^2\,.
\ee
Interestingly, this satisfies, like $(\Delta F)^2$, a negativity property\footnote{Thanks to Laurent Freidel for pointing this out.}:
\bea
(\Delta G)^2 &=& j(j-1) - \frac{1}{4}\left(n^2 - \rho^2 - 4\right) - (m A_j)^2 \\
&=&
j^2 - j - \left(\frac{n}{2}\right)^2 + \frac{\rho^2}{4} + 1 - \left(m \frac{\rho\, n}{4 j(j-1)}\right)^2 \\
&=&
j^2 - \left(\frac{n}{2}\right)^2 - j + 1 - \frac{\rho^2}{4}\left[\left(\frac{m}{j}\frac{n/2}{j-1}\right)^2 - 1\right] \\
&<& 0\quad\mbox{for $j > 1$}\,.
\eea
When applied to coherent states that are subject to the simplicity constraint, the first line gives
\bea
(\Delta G)^2 &=& j(j-1) - \frac{1}{4}\left(n^2 - \rho^2 - 4\right) - (j \tilde{A}_j)^2 \\
&=&
\frac{1}{4} \Big(\rho - \gamma n \Big) \Big(\rho + \frac{n}{\gamma} \Big) - \gamma^2 j + 1
\eea
The expressions are almost the same as in the SU(2) case. By going through analogous arguments
we arrive again at the conditions $\rho = \gamma n$ and $j = n/2$.
These are the same equations as before, but $j = n/2$ refers now to irreps of SU(1,1).
Moreover, $n$ has to be restricted to $n\ge 2$, since $j\ge 1$ for the states in the
Plancherel decomposition.

The quantum area can be obtained from relation \eq{relationNF}.
By squaring this equation and setting the right--hand side equal to the expectation value of the coherent state,
we find that
\be
A^2 = \gamma^2 \b Q\ket = \gamma^2 j(j-1)\,,
\label{squareareaSU(1,1)}
\ee
or
\be
A = \gamma \sqrt{j(j-1)}\,.
\ee
This area is always a non--negative real number, since $j\ge 1$.
Alternatively, we could start from
\be
A^2 = \frac{1}{2} (\star B)^{IJ} (\star B)_{IJ}
\ee
for timelike $N$, and compute
\be
\frac{1}{2} (\star B)^{IJ} (\star B)_{IJ} = \frac{1}{2} (\star B)^{ij} (\star B)_{ij} = \gamma^2 Q\,,\qquad i,j = 0,1,2\,,
\ee
using the simplicity constraints. This gives again eq.\ \eq{squareareaSU(1,1)}.

Clearly, the results for the discrete series are very similar to those for the SU(2) irreps.
This makes sense when we consider the corresponding classical quantities:
the coherent states correspond to spacelike $U$ and timelike $N$, so the area bivector $\star B$ is spacelike,
like in the SU(2) case. Thus, we have described the same classical object in two different gauges:
first for timelike $U$ and then for spacelike $U$.

That we have these two possibilities is important: if the normals $U$ of tetrahedra were always timelike,
we could never have timelike triangles. By allowing also spacelike normals $U$
we permit tetrahedra to contain both spacelike and timelike triangles. The discrete series deals
with the spacelike triangles in such tetrahedra. In the next subsection, we will provide the formalism
for the timelike triangles.

\subsection{Constraints for the continuous series}

When we come to the continuous series, the most important question is: what are the appropriate coherent states?
Since now $Q < 0$, the classical vectors $N$ should be spacelike.
For the continuous series, Perelomov uses the state $|j\,m=0\ket$ and its SU(1,1) transformations \cite{Perelomov}.
This is not the state we are looking for, since it has zero expectation value with regard to $J^3$, $K^1$ and $K^2$. It produces
the zero vector $N = 0$ classically.

What we need is a spacelike vector: eigenstates of $J^3$ are not suited for this, since they lead to vectors $(\pm 1,0,0)$.
This suggests that we use eigenstates of $K^1$ or $K^2$ instead. Such states have been studied by Mukunda \cite{Mukunda_Unitary},
Barut and Phillips \cite{BarutPhillips} and Lindblad and Nagel \cite{LindbladNagel}. We adopt the notation of \cite{LindbladNagel} and write
\be
K^1\, |j\,\lambda\,\sigma\ket = \lambda |j\,\lambda\,\sigma\ket\,.
\ee
The spectrum of $K^1$ is the real line and it is two--fold degenerate. For this reason, eigenstates carry an additional label $\sigma = \pm$
which denotes two orthogonal states with the same eigenvalue $\lambda$. Due to the non--compactness of the $K^1$ subgroup, the states
$|j\,\lambda\,\sigma\ket$ are not normalizable:
\be
\b j\,\lambda'\,\sigma' |j\,\lambda\,\sigma\ket = \delta(\lambda' - \lambda) \delta_{\sigma'\sigma}
\label{orthogonalityjlambda}
\ee
In this respect, they are similar to eigenstates of momentum and a rigorous definition
can be given by using rigged Hilbert space techniques (see \cite{LindbladNagel} for details).

In the following, we will construct coherent states that are based on eigenstates of $K^1$, with the aim of satisfying the
semiclassical conditions \eq{uncertaintyJdiscrete}, \eq{quantumsimplicitydiscrete} and \eq{uncertaintyKdiscrete}.
When doing so, we have to take into account that the states are not normalizable.
This non-normalizability appears at two independent levels:
firstly, we are using eigenstates of $K^1$, so they are not normalizable due to eq.\ \eq{orthogonalityjlambda}.
Secondly, the states $|j\,\lambda\,\sigma\ket$ correspond to states in $\clH_{(\rho,n)}$---let us denote them by
$\left|\Psi^{(\alpha)}_{s\,\lambda\,\sigma}\right\ket$---and these are not normalizable because of
eq.\ \eq{orthogonalitycontinuousseries}.

We deal with this by smearing the states suitably with a Gaussian wavefunction in $\lambda$ and $s$.
The smearing comes with a parameter $\delta$ that we will send to zero in the end\footnote{Compare this with the use of
momentum wavefunctions in scattering amplitudes.}. For the smearing we use the Gaussian
\be
f_{\delta}(s) = \frac{1}{(2\pi)^{1/4} \delta^{1/2}}\,\e^{-s^2 / 4\delta^2}\,.
\ee
The smeared states are defined by
\be
\left|\Psi^{(\alpha)}_{s\,\lambda\,\sigma}(\delta)\right\ket \equiv
\int\limits_0^\infty \d s'\;\sqrt{\mu_\epsilon(s')} f_\delta(s' - s) \int\limits_{-\infty}^{\infty} \d\lambda'\; f_\delta(\lambda' - \lambda)
\left|\Psi^{(\alpha)}_{s'\,\lambda'\,\sigma}\right\ket\,,
\ee
and have norm 1. Their variance in $F$ is
\be
\left(\Delta F\right)^2 =
\left\b \Psi^{(\alpha)}_{s\,\lambda\,\sigma}(\delta)\right| F^i F_i \left|\Psi^{(\alpha)}_{s\,\lambda\,\sigma}(\delta)\right\ket
- \left\b \Psi^{(\alpha)}_{s\,\lambda\,\sigma}(\delta)\right| F^i \left|\Psi^{(\alpha)}_{s\,\lambda\,\sigma}(\delta)\right\ket
\left\b \Psi^{(\alpha)}_{s\,\lambda\,\sigma}(\delta)\right| F_i \left|\Psi^{(\alpha)}_{s\,\lambda\,\sigma}(\delta)\right\ket\,.
\ee
This looks more complicated than it is, since in the limit $\delta\to 0$ the result reduces to a simple analog of eq.\ \eq{variancediscrete}. For instance,
\be
\left\b \Psi^{(\alpha)}_{s\,\lambda\,\sigma}(\delta)\right| K^1 \left|\Psi^{(\alpha)}_{s\,\lambda\,\sigma}(\delta)\right\ket
=
\int\limits_0^\infty \d s'\; f^2_\delta(s' - s) \int\limits_{-\infty}^{\infty} \d\lambda'\; \lambda' f^2_\delta(\lambda' - \lambda)
\;\stackrel{\delta\to 0}{\rightarrow}\; \lambda\,.
\ee
Thus, the variance becomes
\be
\left(\Delta F\right)^2 \;\stackrel{\delta\to 0}{=}\; -s^2 - \frac{1}{4} + \lambda^2
\ee
Clearly, the variance can now take on arbitrary positive and negative values, if we choose $s$ and $\lambda$ suitably.

Condition \eq{uncertaintyJdiscrete} requires us to bring this uncertainty close to zero, so we should choose $\lambda = \sqrt{s^2 + 1/4}$
or $\lambda = s$, for instance. For simplicity, we will select $\lambda = s$ in the following. The difference between these two choices will
not make any difference in the final simplicity constraints.

Our new coherent states are therefore given by
\be
|j\,g\ket_{\mathrm{sp}} \equiv D^j(g) |j\,s\,+\ket\,, \qquad g\in\mathrm{SU(1,1)}\,,
\ee
for states in $\clC^\epsilon_s$,
or
\be
\left|\Psi^{(\alpha)}_{s\,g}\right\ket
\equiv
D^{(\rho,n)}(g) \left|\Psi^{(\alpha)}_{s\,s\,+}\right\ket\,, \qquad g\in\mathrm{SU(1,1)}\,,
\ee
for states in $\clH_{(\rho,n)}$. The choice of $\sigma = +$ is a convention, but not necessary.
The subscript $\mathrm{sp}$ indicates that expectation values of these states correspond to points of the spacelike one--sheeted hyperboloid $\bH_{\mathrm{sp}}$.
Modulo phase and action of $\bZ_2$,
the coherent states can be parametrized by vectors $\v{N}\in \bH_{\mathrm{sp}} \simeq \mathrm{SU(1,1)}/(G_1\otimes\bZ_2)$, i.e.\
\be
|j\,\v{N}\ket \equiv |j\, g(\v{N})\ket_{\mathrm{sp}}\,,
\ee
where $g(\v{N})$ is a representative in the coset defined by $\v{N} \in \bH_{\mathrm{sp}}$.

When dealing with the simplicity constraint \eq{quantumsimplicitydiscrete},
we restrict our attention to the reference vectors $|j\,s\,+\ket$,
since the other states are covered by SU(1,1) covariance. $K^2$, $J^2$, $J^3$ and $K^3$ do not commute with $K^1$
and change the eigenvalue of $\lambda$, so their expectation values vanish.
Hence we only need to analyze the $F^1$ and $G^1$ component of eq.\ \eq{quantumsimplicitydiscrete}.

In another paper \cite{CHrep}, we derive the action of $\mathrm{SL(2,\bC)}$ generators on eigenstates of $K^1$ and obtain
\be
J^1\,|j\,\lambda\,\sigma\ket = (\ldots) |j+1\,\lambda\,\sigma'\ket + \lambda A_j |j\,\lambda\,\sigma\ket + (\ldots) |j-1\,\lambda\,\sigma'\,\ket\,.
\ee
This equation holds for each of the two continuous series in
the decomposition \eq{decompositionSU(1,1)}. The simplicity constraint yields therefore
\be
s - \frac{1}{\gamma}\,s\, A_j = 0
\ee
or
\be
\gamma = A_j = \frac{\rho\, n}{4j(j+1)}\,.
\label{simplicitycont}
\ee
The final step comes from the variance of $G$:
\be
\left(\Delta G\right)^2 \;\stackrel{\delta\to 0}{=}\; j(j+1) - \frac{1}{4}\left(n^2 - \rho^2 - 4\right) + (s\, A_j)^2
\ee
This gives us
\be
\left(\Delta G\right)^2 = \frac{1}{4} \Big(\rho - \gamma n \Big) \Big(\rho + \frac{n}{\gamma} \Big) - \frac{1}{4}\gamma^2 + 1\,.
\ee
The first term vanishes when
\be
\rho = \gamma n
\label{solution1cont}
\ee
or
\be
\rho = -\frac{n}{\gamma}\,.
\label{solution2cont}
\ee
Plugging back \eq{solution1cont} into \eq{simplicitycont} results in a contradiction:
\be
\frac{n^2}{4j(j+1)} = -\frac{n^2}{4(s^2 + 1/4)} = 1\,.
\ee
When inserting \eq{solution2cont} into \eq{simplicitycont}, on the other hand, one obtains
\be
-\frac{\rho^2}{4j(j+1)} = \frac{\rho^2}{4(s^2+1/4)} = 1 \qquad\mbox{or}\qquad \frac{\rho}{2} = -\sqrt{s^2 + 1/4}\,.
\ee
We therefore arrive at the constraints $n = -\gamma\rho$ and $\rho/2 = -\sqrt{s^2 + 1/4}$, which is qualitatively different
from the SU(2) case and the discrete series of SU(1,1)\footnote{Note that we could have also chosen $s = -\rho/2$ and
nevertheless satisfied our semiclassical conditions. We have picked $\rho^2/4 = s^2 + 1/4$, since it leads to a particularly simple
expression for the area spectrum.}. The second condition implies $\rho < -1$ and hence $n > \gamma$.
The area becomes
\be
A = \gamma \sqrt{-\b Q\ket} = \gamma\sqrt{s^2 + 1/4} = -\gamma\frac{\rho}{2} = \frac{n}{2}\,,
\ee
leading to the conclusion that the area of timelike surfaces is quantized!

\section{Summary of constraints}
\label{summaryconstraints}

In this section, we summarize the simplicity constraints and area spectra that we have obtained.
Recall that $U$ is, roughly speaking, the normal vector of a tetrahedron and $N$ is the normal
to a triangle in this tetrahedron. The associated bivector is given by $B = A\,U\wedge N$.
In our analysis, several irreducible representations played a role: the unitary irreps $\clH_{(\rho,n)}$ of $\mathrm{SL(2,\bC)}$,
the unitary irrep $\clD_j$ of SU(2), and irreps $\clD^\pm_j$ and $\clC^\epsilon_s$ of the discrete and continuous series of
SU(1,1).

The following table lists our results for the different choices of $U$ and $N$. The constraints of the EPRL model correspond to the first column.

\renewcommand{\arraystretch}{2.5}
\begin{center}
\begin{tabular}{l@{\qquad}l@{\qquad}l@{\qquad}l}
classical data
& \parbox[t]{2.7cm}{$U = (1,0,0,0)$ \\ $N$ spacelike \\ $\star B$ spacelike}
& \parbox[t]{2.7cm}{$U = (0,0,0,1)$ \\ $N$ timelike \\ $\star B$ spacelike}
& \parbox[t]{2.7cm}{$U = (0,0,0,1)$ \\ $N$ spacelike \\ $\star B$ timelike} \\ [0.9cm] \hline
little group & SU(2) & SU(1,1) & SU(1,1) \\
\parbox{3cm}{relevant \\ irreps} & $\clD_j$ & $\clD^\pm_j$ & $\clC^\epsilon_s$ \\ 
constr.\ on $(\rho,n)$ & $\rho = \gamma n$ & $\rho = \gamma n$ & $n = -\gamma \rho$ \\
constr.\ on irreps & $j = n/2$ & $j = n/2$ & $s^2 + 1/4 = \rho^2/4$ \\
\parbox{3cm}{reference \\ coherent states}
& $|j\, j\ket \in \clD_j$
& $|j\,\pm\!j\ket \in \clD^\pm_j$
& $|j\,s\,+\ket \in \clC^\epsilon_s$ \\
coadjoint orbit & $S^2$ & $\bH_\pm$ & $\bH_{\mathrm{sp}}$ \\
area spectrum & $\gamma\sqrt{j(j+1)}$ & $\gamma\sqrt{j(j-1)}$ & $\gamma\sqrt{s^2 + 1/4} = n/2$ \\
\hline
& \raisebox{0.6cm}[0cm]{$\underbrace{\qquad\qquad\qquad}_{\ds \mathrm{EPRL}}$} & &
\end{tabular}
\end{center}

\section{A spin foam model for general Lorentzian 4--geometries}
\label{spinfoammodel}

We now dispose of constraints for the quantization of spacelike and timelike simple bivectors.
This allows us to define a new spin foam model that extends the EPRL model and describes realistic
Lorentzian 4--geometries, where both spacelike and timelike surfaces appear.

\subsection{Configurations}

The spin foam model is defined on a 4--dimensional simplicial complex $\Delta$ and its dual complex $\Delta^*$.
We denote edges, triangles, tetrahedra and 4--simplices of $\Delta$ by $l$, $t$, $\tau$ and $\sigma$ respectively.
For dual vertices, edges and faces we use $v$, $e$ and $f$ respectively.

Let us start by explaining the way geometrical data are assigned to cells of these complexes.
If we speak in terms of classical variables, we have the following assignments:
there is a normal vector $U$ for each tetrahedron $\tau$, a normal vector $N$ for each triangle $t$ inside a tetrahedron $\tau$
(i.e.\ for each pair $\tau t$), and a bivector $B$ or $\star B$ for each triangle $t$. In terms of the dual $\Delta^*$,
this means that we assign a vector $U_e$ to each edge $e$, a vector $N_{ef}$ to each pair
$ef$, where $f$ contains $e$, and a bivector $B_f$ to each face $f$.
Since we fix the gauge, the value of $U_e$ amounts to a choice between $(1,0,0,0)$ and $(0,0,0,1)$.

At the quantum level, these degrees of freedom translate to the following data:
we choose a little group for each edge $e$, either SU(2) or SU(1,1),
and an irrep of this little group for each pair $ef$, which can be $\clD_j$, $\clD^\pm_j$ or $\clC^\epsilon_s$.
Furthermore, there is an $\mathrm{SL(2,\bC)}$ irrep $(\rho_f,n_f)$ for each face $f$.

These data are subject to the simplicity constraints: the allowed configurations are given by the three columns in
the table of \sec{summaryconstraints}. The first thing to note is that we cannot assign the three possibilities freely to each edge.
The $\mathrm{SL(2,\bC)}$ irrep $(\rho,n)$ is chosen for an entire face $f$, so if one selects the first column and $\rho = \gamma n$ for
one edge $e$ of this face, one cannot pick the third column and $n = -\gamma \rho$ for another edge $e'$ in it.
Classically, this corresponds to the fact that the bivector $\star B$ for a face $f$ is either spacelike or timelike,
and this affects all edges of the face.

An admissible configuration can be obtained by first assigning a representation $(\pm\gamma^{\pm 1} n_f,n_f)$ to each face of the dual complex,
where $\pm 1$ corresponds to spacelike/timelike $\star B$. This is followed by an assignment of a normal vector $U_e$ (and hence little group) to each edge $e$
that is consistent with the given choice of constraints on $(\rho_f,n_f)$. Finally, one selects an irrep of this little group for each pair $ef$, $f\supset e$,
according to the following scheme:
\begin{center}
\begin{tabular}{l@{\quad if\quad}ll@{\quad $\rightarrow$\quad}l}
$\clD_{n_f/2}$, & $\star B$ spacelike, & $U$ timelike & $N$ spacelike, \\
$\clD^{\pm}_{n_f/2}$, & $\star B$ spacelike, & $U$ spacelike & $N$ timelike future/past, \\
$\clC^\epsilon_{\frac{1}{2}\sqrt{n^2_f/\gamma^2-1}}$, & $\star B$ timelike, & $U$ spacelike & $N$ spacelike.
\end{tabular}
\end{center}

A second set of variables comes from the connection. It is implemented as an assignment of $\mathrm{SL(2,\bC)}$ elements $g_e$ to each edge $e$ of the dual complex
$\Delta^*$. More precisely, we split each edge $e$ into two half--edges $ve$ and $ev'$, where $v$ and $v'$ are the endpoints of the edge. To these half--edges,
we assign group elements $g_{ve} = g^{-1}_{ev}$ and $g_{ev'} = g^{-1}_{v'\!e}$.

\subsection{Partition function}

Next we specify the partition function of the model. We first state the definition, and then explain how it arises from BF theory
by imposition of the simplicity constraints.

For the definition, we will use projectors from the $\mathrm{SL(2,\bC)}$ irrep $\clH_{(\rho,n)}$ to the little group irreps
$\clD_j$, $\clD^\pm_j$ and $\clC^\epsilon_s$:
\bea
P_{(\rho,n),j} &=& \sum_{m=-j}^j \left|\Psi_{j\, m}\right\ket \left\b\Psi_{j\, m}\right| \\
P^{\pm}_{(\rho,n),j} &=& \sum_{\pm m=j}^\infty \left|\Psi^\pm_{j\, m}\right\ket \left\b\Psi^\pm_{j\, m}\right| \\
P^\epsilon_{(\rho,n),s}(\delta) &=&
\sum_{\alpha = 1,2} \sum\limits_{\pm m = \epsilon}^\infty \left|\Psi^{(\alpha)}_{s\, m}(\delta)\right\ket \left\b\Psi^{(\alpha)}_{s\, m}(\delta)\right|
\eea
In the case of the continuous series, we project onto the smeared states
\be
\left|\Psi^{(\alpha)}_{s\,m}(\delta)\right\ket \equiv
\int\limits_0^\infty \d s'\;\sqrt{\mu_\epsilon(s')} f_\delta(s' - s)\left|\Psi^{(\alpha)}_{s'\,m}\right\ket\,,
\ee
in accordance with our discussion on normalizability (see \sec{coherencesimplicitySU(1,1)}).
Matrix elements are defined in the limit, where the smearing parameter $\delta$ goes to zero.

The partition function is given by the sum
\be
Z = \int\limits_{\mathrm{SL(2,\bC)}}\!\! \prod_{ev} \d g_{ev} \sum_{n_f} \sum_{\zeta_f=\pm 1} \sum_{U_e}
\;\prod_f (1 + \gamma^{2\zeta_f}) n_f^2\,
A_f\!\left((\zeta_f\gamma^{\zeta_f} n_f,n_f),\zeta_f;U_e;g_{ev}\right)\,.
\label{partitionfunction}
\ee
Let us explain the different elements of this formula:
The $\mathrm{SL(2,\bC)}$ group elements $g_{ev}$ are integrated over with the Haar measure.
For each face, we sum over the positive integers $n_f$.
Furthermore, there is a variable $\zeta_f = \pm 1$ that indicates whether the area bivector of a face is spacelike or timelike respectively.
The normal vector $U_e$ is summed over the two possibilities $(1,0,0,0)$ and $(0,0,0,1)$.
Each face $f$ carries an amplitude $A_f$ which arises from a contraction of parallel transports and projectors from the edges $e$ of the face (see \fig{faceprojectors}):
\be
A_f((\rho,n),\zeta;U_e;g_{ev})
= \lim_{\delta\to 0}\; \tr\left[\prod_{e\subset f} D^{(\rho,n)}(g_{ve}) P_{(\rho,n),\zeta,U_e}(\delta) D^{(\rho,n)}(g_{ev'})\right]
\label{faceamplitude}
\ee
Here, $v$ and $v'$ denote the vertices at the beginning and end of each edge.
The projector $P_{(\rho,n),\zeta,U_e}(\delta)$ depends on $\zeta$ and the $U_e$'s and implements the simplicity constraints on irreps of the little group:
\be
P_{(\rho,n),\zeta,U}(\delta) = \left\{
\begin{array}{l@{\mbox{\qquad if\qquad}}ll}
P_{(\rho,n),n/2}\,, & \zeta = 1\,, & U = (1,0,0,0)\,, \\
\theta(n-2) \left(P^+_{(\rho,n),n/2} + P^-_{(\rho,n),n/2}\right)\,, & \zeta = 1\,, & U = (0,0,0,1)\,, \\
\left(1-\theta(\gamma-n)\right) P^\epsilon_{(\rho,n),\frac{1}{2}\sqrt{n^2/\gamma^2-1}}(\delta)\,, & \zeta = -1\,, & U = (0,0,0,1)\,, \\
0\,, & \zeta = -1\,, & U = (1,0,0,0)\,.
\end{array}
\right.
\ee

\psfrag{P}{$P$}
\psfrag{f}{$f$}
\psfrag{e}{$e$}
\psfrag{v}{$v$}
\psfrag{v'}{$v'$}
\pic{faceprojectors}{Graphical representation of face amplitude $A_f$:
lines stand for representation matrices $D^{(\rho,n)}(g_{ve})$ and pairs of dots symbolize projectors $P_{(\rho,n),\zeta,U_e}$.}{4cm}{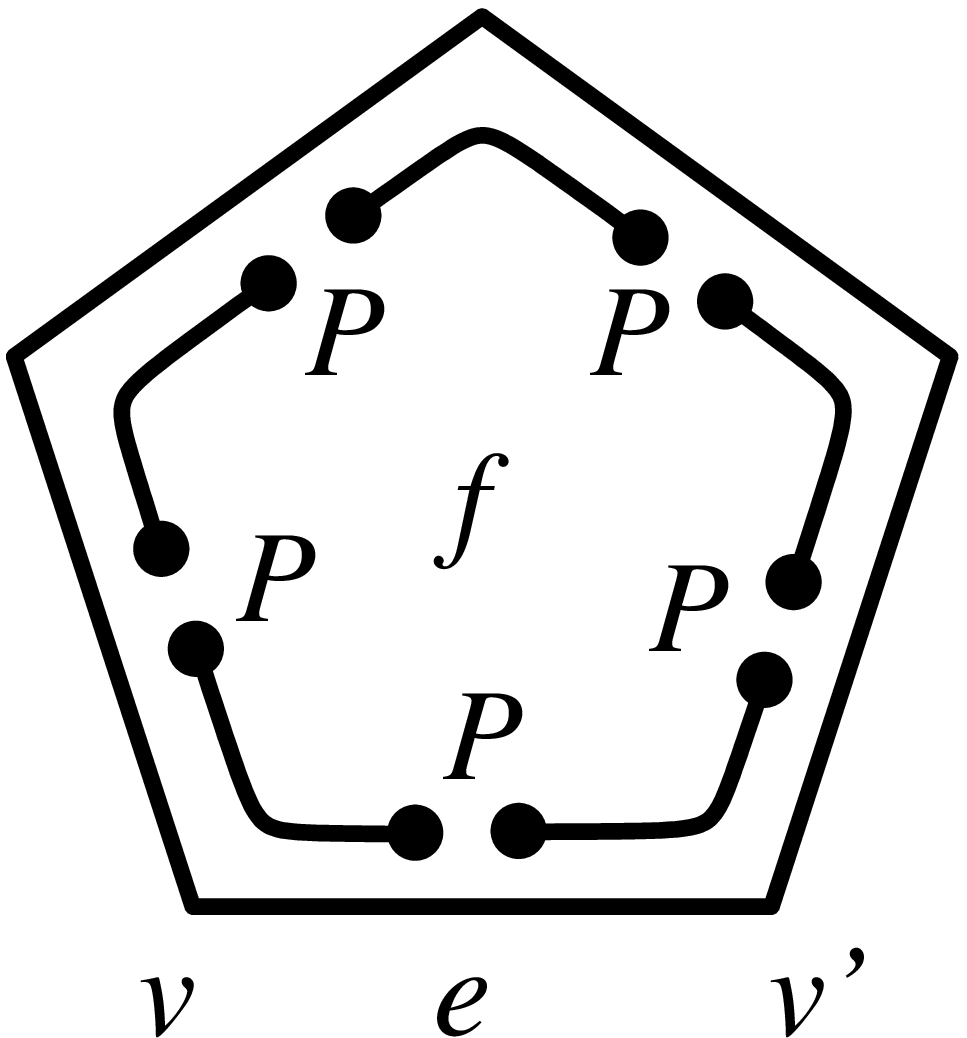}

Formulas \eq{partitionfunction} and \eq{faceamplitude} are the result of imposing simplicity constraints on quantum BF theory.
The spin foam model of $\mathrm{SL(2,\bC)}$ BF theory has the partition function
\be
Z_{\mathrm{BF}} = \int\limits_{\mathrm{SL(2,\bC)}}\!\! \prod_{ev} \d g_{ev} \sum_{n_f} \int\limits_{-\infty}^{\infty} \prod_f \d\rho_f
\,\prod_f (n^2_f + \rho^2_f)\,
A_f\left((\rho_f,n_f);g_{ev}\right)\,,
\label{partitionfunctionBF}
\ee
where
\be
A_f((\rho,n);g_{ev})
= \tr\left[\prod_{e\subset f} D^{(\rho,n)}(g_{ve}) \mathbbm{1}_{(\rho,n)} D^{(\rho,n)}(g_{ev'})\right]\,.
\label{faceamplitudeBF}
\ee
The transition to \eq{partitionfunction} consists of the following steps:
first we impose the simplicity constraint on $(\rho_f,n_f)$, so we introduce the variable
$\zeta_f = \pm 1$ for encoding space-- or timelikeness of a face and set $(\zeta_f\gamma^{\zeta_f}n_f,n_f)$.
Then, we include a sum over normal vectors $U_e$ for each edge $e$, taking the values $(1,0,0,0)$
or $(0,0,0,1)$. The identity operators $\mathbbm{1}_{(\rho,n)}$ in the contraction \eq{faceamplitudeBF}
decompose into irreps of SU(2) or SU(1,1) according to formulas \eq{completenessrelationSU(2)} and \eq{completenessrelationSU(1,1)}.
Depending on the value of $\zeta_f$ and $U_e$ we impose simplicity constraints on these irreps.
This is done by the replacement
\be
\mathbbm{1}_{(\rho,n)}\quad\longrightarrow\quad P_{(\rho,n),\zeta,U}(\delta)\,.
\ee

\subsection{Coherent state vertex amplitude}

There are several equivalent ways in which a spin foam sum
can be written down. One possibility is to express the total amplitude as a product of vertex amplitudes.
Each vertex amplitude can be interpreted as the amplitude for a single 4--simplex.
Furthermore, there are different definitions of the vertex amplitude, depending on the choice of boundary
data for the 4--simplex.
In references \cite{CFquantumgeometry,BarrettasymptoticsEuclidean,BarrettasymptoticsLorentzian}
vertex amplitudes were defined in terms of coherent states that encode the geometry of boundary tetrahedra.

Below we specify this type of vertex amplitude for the present model. For this we use the coherent states
\[
\begin{tabular}{l}
$|j\,g\ket \equiv D^j(g)|j\,j\ket\in \clD_j\,,$ \\
$|j\,g\ket_{\pm} \equiv D^j(g) |j \pm\! j\ket\in \clD^\pm_j\,,$ \\
$|j\,g\ket_{\mathrm{sp}} \equiv D^j(g) |j\,s\,+\ket\in \clC^\epsilon_s\,,$
\end{tabular}
\]
which already appeared in the derivation of the simplicity constraint in \sec{coherencesimplicitySU(2)}
and \sec{coherencesimplicitySU(1,1)}. More precisely, we work with the corresponding coherent states
in the Hilbert space $\clH_{(\rho,n)}$, i.e.\
\[
\begin{tabular}{l@{\qquad}l}
$\left|\Psi_{j\,g}\right\ket \equiv D^{(\rho,n)}(g) \left|\Psi_{j\,j}\right\ket\,,$ & $g\in \mathrm{SU(2)}\,,$ \\
$\left|\Psi^\pm_{j\,g}\right\ket \equiv D^{(\rho,n)}(g) \left|\Psi^\pm_{j\,\pm\! j}\right\ket\,,$ & $g\in \mathrm{SU(1,1)}\,,$ \\
$\left|\Psi^{(\alpha)}_{s\,g}\right\ket \equiv D^{(\rho,n)}(g) \left|\Psi^{(\alpha)}_{s\,s\,+}\right\ket\,,$ & $g\in \mathrm{SU(1,1)}\,.$
\end{tabular}
\]
With the parametrization in terms of quotient spaces (see appendix \ref{parametrization}) one can write such states in a uniform way as\footnote{To be precise,
there is the additional label $\alpha = 1, 2$ for states of the continuous series. For simplicity, we omit this label in the following formula.}
\be
\left|\Psi_{j\v{N}}\right\ket\,,\quad \mbox{where $\v{N} \in S^2$, $\bH_+$, $\bH_-$ or $\bH_{\mathrm{sp}}$.}
\label{coherentstateuniformnotation}
\ee
Consider a single vertex $v$ of the dual complex, and number the edges $a = 1,\ldots 5$.
Accordingly, faces $f$ are labelled by unordered pairs $ab$ and pairs $ef$ of edges and faces are encoded by
ordered pairs $ab$. The vertex amplitude is given by the product
\be
A_v = \int\limits_{\mathrm{SL(2,\bC)}} \prod_a \d g_a\; \prod_{a < b} A_{ab}\,,
\ee
where
\be
A_{ab} = \left\b \Psi_{j_{ab}\v{N}_{ab}}\right| D^{(\rho_{ab},n_{ab})}(g_a g^{-1}_b) \left| \Psi_{j_{ab}\v{N}_{ba}}\right\ket
\label{coherentstatevertexamplitude}
\ee
The variables $g_a \in \mathrm{SL(2,\bC)}$ represent the connection on half--edges adjacent to the vertex.
$\rho_{ab}$ and $n_{ab}$ are subject to the constraint $\rho_{ab} = \pm\gamma^{\pm 1} n_{ab}$ for spacelike/timelike
triangles. The coherent states $| \Psi_{j_{ab}\v{N}_{ba}}\ket$ are states of the type \eq{coherentstateuniformnotation}
and the allowed choices for $j_{ab}$ and $\v{N}_{ab}$ follow from the table in \sec{summaryconstraints}.

\section{Discussion}

Let us summarize our results.
We derived simplicity constraints for the quantization of general Lorentzian 4--geometries.
The method for this derivation was based on coherent states.

The constraints operate at two levels\footnote{See also the table in \sec{summaryconstraints}.}: on irreps of $\mathrm{SL(2,\bC)}$
and on irreps of subgroups of $\mathrm{SL(2,\bC)}$.
Firstly, there arise two possible restrictions on unitary irreps of $\mathrm{SL(2,\bC)}$,
one for spacelike triangles and one for timelike triangles.
Secondly, we have constraints on unitary irreps of SU(2) and SU(1,1).
These irreps appear in the decomposition of irreps of $\mathrm{SL(2,\bC)}$
and encode the geometry of triangles. Irreps of SU(2) correspond to triangles
in tetrahedra with a timelike normal, while irreps of SU(1,1) refer to triangles in
tetrahedra with a spacelike normal. In the latter case, spacelike and timelike triangles
are described by states in the discrete and continuous series of SU(1,1) respectively\footnote{In the context of 3d gravity,
the relation between geometry and irreps of SU(1,1) was analyzed in \cite{Davids} and \cite{Freidel_Spectra}.}.

The constraints of the EPRL model were reproduced in the special case, when
tetrahedra have only timelike normals.
In all cases, we obtained a discrete area spectrum.
Thus, the discreteness of area in loop quantum gravity
is extended to timelike surfaces.

The derivation of the constraints rested on the idea that coherent states should mimic classical simple bivectors
as closely as possible. For this reason, we constructed coherent states with the following properties:
\begin{enumerate}
\item Expectation values of bivectors satisfy the classical simplicity constraint.
\item The uncertainty in bivectors is minimal.
\end{enumerate}
It turned out that such states can only exist in certain irreps of $\mathrm{SL(2,\bC)}$, SU(2) and SU(1,1),
and this is what determined the conditions on irreps.

Based on these constraints, we defined a new spin foam model that provides a quantization of
general Lorentzian 4--geometries. The geometries are general in the sense that they admit
both spacelike and timelike triangles, whereas in the EPRL model only spacelike triangles were permitted.
The timelike triangles may appear, in particular, in the boundary of the simplicial complex, so one
can now consider regions with timelike boundaries.

In regard to future work, there is a whole range of results that apply to the EPRL model
and could be extended to the new model: for instance,
the semiclassical limit, given by the asymptotics in the large area limit \cite{CFsemiclassical,BarrettasymptoticsEuclidean,BarrettasymptoticsLorentzian},
the derivation of graviton propagators \cite{gravitonpropagatornewmodel} and the description of intertwiners in terms of tetrahedra \cite{CFquantumgeometry}.
This will require the use of new techniques, since SU(2) will be replaced by the less familiar SU(1,1)
and its representation theory.

Since we were dealing with timelike surfaces and their area spectrum, our results should be compared to previous work
in this direction, e.g.\ \cite{PerezRovellitimelike} and \cite{AlexandrovVassilevich,AlexandrovKadar}.

So far our method has been only applied to the Lorentzian case, so it would be interesting to see
what it will give in the Euclidean regime.
For this, one will have to use coherent states in the SU(2) decomposition of SU(2)$\times$SU(2) irreps
and compute their expectation values. Will this reproduce the Euclidean EPRL model?
If it does, this will shed further light on the relation between the Euclidean FK and EPRL model,
since then the construction of the two models can be directly compared.

A point that we left open is the definition of the spin foam model as an integral over coherent states.
We did specify a vertex amplitude with coherent states as boundary data, but we did not use it to define
the spin foam sum as a whole. For this, we would need a completeness relation for all three types of coherent states
that appear in our construction. For SU(2) and for the discrete series of SU(1,1) such completeness relations were
given by Perelomov \cite{Perelomov}. For the continuous series, however, we have introduced a new type of coherent state
and we have not yet provided an associated completeness relation.

\begin{acknowledgments}
We thank Laurent Freidel for discussions. Research at Perimeter Institute is supported by the Government of Canada through Industry Canada and by the Province of Ontario
through the Ministry of Research \& Innovation.
\end{acknowledgments}

\begin{appendix}

\section{Parametrization of coherent states}
\label{parametrization}

In this appendix, we explain the parametrization of coherent states in terms of hyperboloids.
Consider first the discrete series and the coherent states
\be
|j\,g\ket_\pm \equiv D^j(g) |j \pm\! j\ket\in \clD^\pm_j\,.
\ee
Up to phase, the states are determined by cosets in $\mathrm{SU(1,1)} / \mathrm{U(1)}$.
Elements $g\in\mathrm{SU(1,1)}$ can be parametrized by
\be
g = \e^{-\irm \varphi J^3}\, \e^{-\irm u K^1}\, \e^{\irm \psi J^3}\,,\qquad -\pi < \varphi \le \pi\,,\quad -2\pi < \psi \le 2\pi\,,\quad 0 \le u < \infty\,,
\ee
so the cosets have representatives
\be
g = \e^{-\irm \varphi J^3}\, \e^{-\irm u K^1}\,.
\label{representative+}
\ee
The parameters $\varphi$ and $u$ coordinatize the upper/lower hyperboloid via
\be
\v{N} = \pm (\cosh u,\sinh u \sin\varphi,\sinh u \cos\varphi)\,,\qquad \v{N}^2 = 1\,,
\label{isomorphismHpm}
\ee
giving the isomorphism $\mathrm{SU(1,1)} / \mathrm{U(1)} \simeq \mathbbm{H}_\pm$.
Thus, the representatives \eq{representative+} can be expressed as functions $g(\v{N})$ of $\v{N}\in\mathbbm{H}_\pm$,
and the coherent states may be defined by
\be
|j\,\v{N}\ket \equiv |j\,g(\v{N})\ket\,.
\ee
Formula \eq{isomorphismHpm} was chosen such that
\be
\b j\,\v{N}| \v{F} |j\,\v{N}\ket = j\v{N}\,.
\ee

For the continuous series, we employ coherent states
\be
|j\,g\ket_{\mathrm{sp}} \equiv D^j(g) |j\,s\,+\ket\in \clC^\epsilon_s\,.
\ee
In this case, we consider the quotient $\mathrm{SU(1,1)} / (G_1\otimes\bZ_2)$, where $G_1$ is
the one--parameter subgroup generated by $K_1$.
Now, it is convenient to choose the following parametrization \cite{Lindblad}:
\be
g = \e^{-\irm \varphi J^3}\, \e^{-\irm t K^2}\, \e^{\irm u K^1}\,,\qquad -2\pi < \varphi \le 2\pi\,,\quad-\infty < t, u < \infty\,.
\ee
Representatives of cosets are given by elements
\be
g = \e^{-\irm \varphi J^3} \e^{-\irm t K^2}\,,\qquad -\pi < \varphi \le \pi\,,\quad-\infty < t < \infty\,.
\label{representative_s}
\ee
Using
\be
\v{N} = (-\sinh t,\cosh t\cos\varphi,\cosh t\sin\varphi)\,,\qquad \v{N}^2 = -1\,,
\label{isomorphismHs}
\ee
we parametrize the single--sheeted hyperboloid $\bH_{\mathrm{sp}}$ by $(\varphi,t)$, and
it follows that $\mathrm{SU(1,1)} / (G_1\otimes\bZ_2) \simeq \bH_{\mathrm{sp}}$.
Hence we write representatives \eq{representative_s} as functions $g(\v{N})$ of $\v{N}\in\bH_{\mathrm{sp}}$.
The corresponding coherent states are specified by
\be
|j\,\v{N}\ket \equiv |j\,g(\v{N})\ket_{\mathrm{sp}}\,,
\ee
where
\be
\b j\,\v{N}| \v{F} |j\,\v{N}\ket = s\v{N}\,.
\ee

\end{appendix}

\end{document}